\documentclass[prd,preprint,superscriptaddress,showpacs,nofootinbib]{revtex4}
\newcommand{\ben}{\begin{displaymath}}
\newcommand{\een}{\end{displaymath}}
\newcommand{\be}{\begin{equation}}
\newcommand{\ee}{\end{equation}}
\newcommand{\bea}{\begin{eqnarray}}
\newcommand{\eea}{\end{eqnarray}}

\begin{document}
\preprint{MKPH-T-03-17}
%\draft
\title{Is the strong-interaction proton-proton scattering length
renormalization
scale dependent in effective field theory?}
\author{J.~Gegelia\footnote{e-mail: gegelia@kph.uni-mainz.de}}
\thanks{Alexander von Humboldt Research Fellow}
\affiliation{\it Institut f\"ur Kernphysik, Johannes
Gutenberg-Universit\"at, J.J. Becherweg 45, \\ D-55099 Mainz,
Germany}
\affiliation{Department of Physics, Flinders University, Bedford
park, S.A. 5042, Australia}
\affiliation{High Energy Physics Institute
of TSU, University str. 9, Tbilisi 380086, Georgia}

\begin{abstract}
It is shown that the strong-interaction ${ }^1S_0$ proton-proton
scattering length in very low-energy effective field theory does
not depend on the renormalization scale, if the electromagnetic
interaction is "switched off" consistently.
%
%Shortened title:

%Proton-proton scattering length in effective field theory

\end{abstract}

%Nucleon-nucleon interactions (including antinucleons, deuterons,
%          etc.)

\medskip
\medskip
\date{\today}
\maketitle

Over the past decade Weinberg's papers on describing nuclear
forces using chiral Lagrangians
\cite{Weinberg:1990rz,Weinberg:1991um} have triggered an intensive
activity (see, {\it e.g.}, refs.
\cite{Bedaque:2000kn},\cite{Beane:2000fx},\cite{Meissner:2003ut}
and references therein). For processes involving more than one
nucleon Weinberg suggested to apply the power counting to the
effective potentials. The transition amplitudes are then obtained
by solving the Lippmann-Schwinger equation (or the Schr\" odinger
equation). This approach
%as well as a new power counting (applicable directly to
%amplitudes) suggested in ref. \cite{Kaplan:1998tg}
has been
applied to various problems involving two and three nucleons.

In this work we address the dependence of the strong-interaction
proton-proton scattering length on the renormalization mass
parameter encountered in ref. \cite{Kong:1999sx}. The discussion
below closely follows the paper by X.~Kong and F.~Ravndal
\cite{Kong:1999sx} and the author's PhD thesis \cite{thesis}.
Similar considerations have recently been presented independently
in refs. \cite{Rusetsky:2002kv,Gasser:2003hk}.

In order to describe proton-proton scattering at very low
energies, one can integrate out all particles except protons and
photons. The lowest-order strong interaction part of the effective
non-relativistic Lagrangian for protons in the spin-singlet
channel reads \cite{Kong:1999sx}

\be
    {\cal L}_0 = \psi^\dagger\left(i\partial_0 +
{{{\bf {\nabla}}}^2\over 2M}\right)\psi
             - {C_0\over 2}(\psi\sigma_2\psi )(\psi\sigma_2 \psi
             )^\dagger ,
\label{Leff} \ee where $\psi $ is the two-component field of the
proton, $M$ the mass of the proton, $C_0$ a coupling constant and
$\sigma_2$ a Pauli matrix. This Lagrangian corresponds to the
singular potential $C_0\delta({\bf r})$  which affects
interactions only in the $S$ wave. On top of eq. (\ref{Leff}) one
also needs to include the static Coulomb repulsion between
protons. The effective strength of this repulsion is
$\eta(p)\equiv \eta = \alpha M/2p$ where $p$ is the magnitude of
the CM momentum of the protons and $\alpha =1/137$ is the fine
structure constant. For small $p$, $\eta $ is large and hence the
Coulomb repulsion becomes strong. The scattering problem for both
the Coulomb repulsion and the singular strong-interaction
potential of eq. (\ref{Leff}) can be solved simultaneously using
the well-established formalism based upon the exact solutions of
the Schr\"odinger equation in the Coulomb potential \cite{Coulomb}.

In a partial wave expansion of the full scattering amplitude
\cite{GW}, the total phase shifts $\delta_l$ can be written as
$\sigma_\ell + \delta_\ell^C$, where $\sigma_\ell$ are the pure
Coulombic phase shifts.
% and $\delta_\ell^C$ are the shifts due to
%the strong interactions.
For $pp$-$S$-wave scattering,
$\delta_{pp}^C$ is related to the corresponding (modified) strong
amplitude $T_{SC}(p)$ by the standard partial wave expression
\be
    p\,(\cot\delta_{pp}^C - i) = - {4\pi\over M} {e^{2i\sigma_0}\over
T_{SC}(p)}. \label{Ccot} \ee Note that $\delta_{pp}^C$ besides
pure strong-interaction effects still contains remnants of the
electromagnetic interaction.
%The superscript $C$ is a reminder of
%the fact that $\delta_{pp}^C$ is not the strong phase shift in the
%absence of Coulomb effects. (Instead it is the strong phase shift
%modified by the Coulomb potential).
It is only the Coulomb repulsion between the protons in the
initial and final states that has been removed at this stage.

It is well known that  $\cot\delta_{pp}^C$ in eq.~(\ref{Ccot})
does not have a regular effective range expansion. Rather one
finds \cite{Bethe}
\be
    p\left[C_\eta^2\left(\cot\delta_{pp}^C - i\right) + 2\eta H(\eta)\right]
    = - {1\over a_{pp}^C} + {1\over 2}r_0\,p^2 + \ldots,
\label{CERE}
\ee
where
\be
     C_\eta^2 = {2\pi\eta\over e^{2\pi\eta} - 1}
\label{Sommer} \ee is the Sommerfeld factor \cite{Coulomb},
$a_{pp}^C$ and $r_0$ are the $S$-wave Coulomb-modified scattering
length and effective range, respectively. They arise after
removing the part of the amplitude described by the complex
function \cite{Haeringen}
\be
     H(\eta) = \psi(i\eta) + {1\over 2i\eta} - \ln(i\eta)
\label{Hfun} \ee representing Coulomb effects at short distances.
Here, the $\psi$ function is the logarithmic derivative of the
$\Gamma $ function. The imaginary part of eq.~(\ref{Hfun}) cancels
the term $\sim i$ in eq.~(\ref{CERE}). The real part defines the
function  $h(\eta) = \mbox{Re}\psi(i\eta) - \ln\eta$ which is more
suitable for a phenomenological analysis \cite{BJ}.

The proton-proton scattering amplitude can be calculated from the
effective Lagrangian of eq.~(\ref{Leff}) (plus the Coulomb term).
It takes the form \cite{Kong:1999sx}:
\be
     T_{SC}(p) =  C_{\eta}^2{C_0\,e^{2i\sigma_0}\over 1 -
     C_0\,J_0(p)},
\label{TSC}
\ee
where
\be
      J_0(p) = M\!\int\!{d^3 k\over (2\pi)^3} {2\pi\eta(k)\over e^{2\pi\eta(k)}
- 1} {1\over p^2 - k^2 + i\epsilon}. \label{Cbubble} \ee When this
result for the scattering amplitude is used in eqs.~(\ref{Ccot})
and (\ref{CERE}), we see that both the phase shift $\sigma_0$ and
the Sommerfeld factor $C^2_\eta$ cancel out. We are thus only left
with the evaluation of eq.~(\ref{Cbubble}) which can be done using
the power divergent subtraction (PDS) scheme of
ref.~\cite{Kaplan:1998tg} in $d = 3 - \epsilon $ dimensions introducing
a renormalization mass $\mu$. An ultraviolet divergence shows up
as an $1/\epsilon$ pole in the integral. This will be cancelled by
counterterms which renormalize the coupling $C_0$ in
eq.~(\ref{TSC}) to $C_0(\mu)$. As a result, the finite part of the
dressed bubble, eq.~(\ref{Cbubble}), is found to be
\cite{Kong:1999sx}
\be
      J_0^{finite}(p) = {\alpha M^2\over 4\pi}\left
[ \ln{\mu\sqrt{\pi}\over\alpha M}
                  + 1 - {3\over 2}C_E - H(\eta) \right] - {\mu M\over
                  4\pi},
\label{J0fin} \ee where $C_E = 0.5772\cdots$ is Euler's constant.
The last term of eq.~(\ref{J0fin}) is the contribution from the
special PDS pole in $d=2$ dimensions. We now see that also the
function $H(\eta)$ cancels out in eq.~(\ref{CERE}). At this order
in the effective theory there is no contribution to the effective
range $r_0$.
If one defines the strong scattering length (with the
Coulomb interaction switched off) as \cite{Kong:1999sx}
\be
    {1\over a_{pp}} = {4\pi\over MC_0(\mu)} +
    \mu,
\label{app} \ee then from eqs.~(\ref{CERE}) and (\ref{TSC}) it can
be expressed in terms of the measured scattering length $a_{pp}^C$
as
\be
    {1\over a_{pp}} =  {1\over a_{pp}^C} + \alpha M
    \left[\ln{\mu\sqrt{\pi}\over\alpha M} + 1 - {3\over
    2}C_E\right].
\label{Capp} \ee As is seen from eq.~(\ref{Capp}), $a_{pp}$
depends on $\mu $. It has been argued in ref. \cite{Kong:1999sx}
that the strong scattering length $a_{pp}$ is not a physical
quantity as it can not be measured directly and thus in general it
can depend on the renormalization point $\mu$. However, this
explanation does not seem very satisfactory if $a_{pp}$ is indeed
understood as a scattering length for the strong interaction when
the Coulomb interaction is switched off. It is completely true
that $a_{pp}$ is not measurable experimentally but from a
theoretical point of view it still is a physical quantity.
As it is clear from the analyses of ref. \cite{Kong:1999sx} the sum
of $\delta$ function and the Coulomb potential is renormalizable.
The $\delta$ function potential is also renormalizable.
Hence as one could consider these two potetials themselves as independent
'models' (without higher order corrections of effective field theory),
all physical quantities of both 'models' should be
renormalization  point independent. Consequently one {\it can not}
expect that the dependence of the strong scattering length on the
renormalization point can be canceled by contributions of
higher-order terms in the potential generated by effective field
theory as suggested in ref. \cite{Kong:1999sx}. The origin of the
$\mu$ dependence of $a_{pp}$ is the $\alpha$ dependence of the
running of $C_0(\mu )$ in eq.~(\ref{app}). In order to define the
strong scattering length consistently as the quantity of the
theory with the electromagnetic interaction being switched off, we
should also put $\alpha =0$ in the running of $C_0(\mu )$.

It is straightforward to calculate the running of the renormalized
coupling constant $C_0(\mu )$ using the results of ref. [6]:
\be
C_0(\mu )={C_0\left( \mu_0\right)\over 1-C_0\left( \mu_0\right)
\left\{ -{\alpha M^2\over 4\pi}\ln {\mu\over \mu_0}+{M\over
4\pi}\left( \mu-\mu_0\right)\right\}}. \label{c0mu} \ee Setting
$\alpha =0$ in eq. (\ref{c0mu}) yields
\be
\tilde C_0(\mu )={\tilde C_0\left( \mu_0\right)\over 1-\tilde C_0
\left( \mu_0\right) {M\over 4\pi}\left( \mu-\mu_0\right)}
\label{tildec0mu} \ee Note that the value of the strong coupling
constant $\tilde C_0\left( \mu \right)$ for $\mu =\mu_0$ when the
Coulomb interaction is switched off does not coincide with
$C_0\left( \mu\right)$. Defining
\be
    {1\over a_{pp}} =-p \left[ C_\eta^2\left( \cot\delta^C_{pp}-i\right)
    +2 \eta H(\eta)\right]|_{p=0,\alpha =0}= {4\pi\over M\tilde C_0(\mu)} +
    \mu,
\label{appnew} \ee one obtains the strong scattering length which
does not depend on the renormalization parameter $\mu $. The
numerical value of $\tilde C_0(\mu_0)$ for some fixed $\mu_0$ has
to be given as an input, as it {\it can not} be calculated from
$C_0\left( \mu\right)$ within the given effective field theory.
This is in agreement with the result of ref.~\cite{Sauer:jh} that the
${ }^1S_0$ $pp$ scattering amplitude can not be devided into strong
and electromagnetic parts in a model independent way.

%This last fact is a characteristic feature for any quantum field
%theory with two (or more) coupling constants.

Note that if we consider both protons and neutrons with an isospin
invariant contact interaction in the ${ }^1S_0$ partial wave then
at given order of accuracy $\tilde C_0(\mu)$ exactly coincides with
the renormalised coupling of $pn$ and $nn$ contact interactions, and
consequently $a_{pp}$ coinsides with $a_{pn}=a_{nn}$. This pins down
$\tilde C_0(\mu)$. Unfortunately this is only the manifastation of the
isospin symmetry which has been taken as an input. Given $\tilde C_0(\mu)$
as an input (fixed through $a_{nn}$) one can not calculate $C_0(\mu )$
(and consequently $a^C_{pp}$) within given effective theory.

\medskip
\medskip

One can also include the next-to-leading order correction to the effective
Lagrangian (\ref{Leff}) which for $S$-wave channel reads \cite{Kong:1999sx}:

\begin{equation}
{C_2\over 16} \ (\psi\sigma_2 ( \stackrel{\rightarrow}\nabla-\stackrel
{\leftarrow}\nabla)^2
\psi )(\psi\sigma_2 \psi
             )^\dagger +h.c.
\label{Leff2}
\end{equation}
Taking into account the contribution of ${\cal L}_2$ into $pp$
potential one obtains the amplitude (in dimensional
regularization) in exact analogy to the case of the contact
interactions plus one pion exchange potential of the
ref.~\cite{Kaplan:1996xu}:
\be
T_{SC}(p) =  {\psi(0)^2\over \frac{1}{C_0+C_2 p^2} -
     G_E(0,0)},
\label{TSC2ksw} \ee where $\psi(0)$ is Coulomb wave function at
the origin and $G_E(0,0)$ is the coordinate-space propagator from
the origin to the origin in the presence of Coulomb potential.
Substituting the values of these two quantities \cite{Kong:1999sx}
we obtain
\be
     T_{SC}(p) =  C_{\eta}^2{e^{2i\sigma_0}\over \frac{1}{C_0+C_2 p^2} -
     J_0(p)}.
\label{TSC2} \ee The model is no longer renormalizable, i.e. not
all divergences can be absorbed in available parameters, but if
one expands in powers of $C_2$ and keeps only the zeroth and first
order terms (in $C_2$), then all divergences can be absorbed in
$C_0$ and $C_2$. The renormalization is performed analogously to
the case without Coulomb interaction
\cite{Gegelia:1998iu,Gegelia:1999ja}, i.e. one expands

\be
     T_{SC}(p) =  C_{\eta}^2{e^{2i\sigma_0}\over \frac{1}{C_0} -
     J_0(p)}+C_{\eta}^2{e^{2i\sigma_0} p^2 \ C_2\over C_0^2 \left( \frac{1}{C_0} -
     J_0(p)\right)^2}+O\left( p^4 \ C_2^2\right)
\label{TSC2exp} \ee and absorbs the divergences of $J_0(p)$ by
counterterms which renormalize $C_0$ leading to running coupling
of eq.~(\ref{c0mu}). The remaining divergences contained in
$C_0^2$ factor in the denominator of the second term in
eq.~(\ref{TSC2exp}) is absorbed into the renormalization of $C_2$
by demanding
\begin{equation}
\frac{C_2}{C_0^2}=\frac{C_2(\mu)}{C_0(\mu)^2}. \label{c2renorm}
\end{equation}
As the left-hand side of the eq.~(\ref{c2renorm}) is the ratio of
bare couplings it does not depend on $\mu$, hence the right-hand
side does not depend on $\mu$ either. Writing for some fixed
$\mu_0$
\begin{equation}
\frac{C_2(\mu)}{C_0^2(\mu)}=\frac{C_2(\mu_0)}{C_0(\mu_0)^2},
\label{c2renormmu0}
\end{equation}
solving eq.~(\ref{c2renormmu0}) for $C_2(\mu)$ and taking the
eq.~(\ref{c0mu}) into account one obtains the following
expression:

\be
C_2(\mu )={C_2\left( \mu_0\right)\over \left( 1-C_0\left(
\mu_0\right) \left\{ -{\alpha M^2\over 4\pi}\ln {\mu\over
\mu_0}+{M\over 4\pi}\left( \mu-\mu_0\right)\right\}\right)^2}.
\label{c2mu} \ee

Analogously to $C_0(\mu)$, the running of $C_2(\mu)$ depends on $\alpha$.
Therefore to obtain the effective theory with electromagnetic interaction
switched off, one should put the fine structure constant equal to zero in
the running of $C_2$ as well. This will lead to running coupling
$\tilde C_2(\mu)$
which can not be calculated from $C_2(\mu)$. In fact the running of all
couplings
of low energy effective field theory of strong and electromagnetic
interactions depends on $\alpha$. This dependence has to be
switched off together with the explicit $\alpha $-dependence if one considers
the quantities of the theory with electromagnetic interaction
switched off.

The underlying 'fundamental theory' of strong interactions, QCD,
is most likely an effective theory itself \cite{Weinberg:mt}. The only
parameters of this theory which we can meaningfully interpret are the
renormalized, running parameters. Therefore the electromagnetic and
strong interaction contributions in physical quantities can not be
unambigously separated in this theory either (for detailed analyses
see ref.~\cite{Gasser:2003hk}). The unambigous separation of the
electromagnetic and strong
interaction contributions in physical quantities would be possible in a truly
fundamental, nonperturbatively finite theory (string theory, M-theory?).
On the other hand if we consider
the renormalized parameters of both QCD and QCD+electromagnetic interaction
as an input, one can calculate (at least in principle) the low energy
constants of effective field theories with and without electromagnetic
interaction using some non-perturbative technique like lattice
calculations.

\medskip

\medskip

In conclusion, we have considered ${ }^1S_0$ proton-proton
scattering at very low energies in the framework of effective
field theory, where all degrees of freedom except the proton and
the photon are integrated out. We have argued that the dependence
of the strong proton-proton scattering length  (with the Coulomb
interaction switched off) on the renormalization mass parameter
occurs only if the Coulomb interaction is not completely switched
off. To consider quantities entirely due to the strong
interactions, one should also turn off the Coulomb interaction in
the running of the strong interaction coupling. Doing so generates
the strong interaction proton-proton scattering length which does
not depend on the renormalization mass parameter.

%It is  expected from general
%consideratins that as while
%experimentally this quantity is not measurable (Coulomb interaction is always
%there) theoretically it is a well defined physical quantity.

\medskip
\medskip
\medskip

{\bf Acknowledgements}

\medskip

%\section{acknowledgements}
The author would like to thank B.~Blankleider, S.~Scherer and
A.~Rusetsky for useful discussions and S.~Scherer for numerous
comments on the manuscript.

The support of the Alexander von Humboldt Foundation is
acknowledged.

\medskip
\medskip
\medskip

{\bf Appendix A}

\medskip

In this appendix we illustrate the solution of the $\mu$ -
dependence problem by means of a simple toy model. Our model is
analogous to $pp$ scattering in the sense that a ``physical
quantity'' exhibits a renormalization-scale dependence when one of
the ``coupling constants'' is put equal to zero but is not
simultaneously switched off in the running of the second coupling
constant.

Suppose we have some ``physical quantities''

\begin{eqnarray}
y_1(p)&=&C_0 \ \left[ a(p)+\alpha \ b(p)\right], \label{2} \\
y_2(p)&=&\alpha \ d_1(p)+C_0 \ d_2(p), \label{3}
\end{eqnarray}
where $\alpha $ and $C_0$ are ``coupling constants'' and $a(p)$,
$b(p)$, $d_1(p)$ and $d_2(p)$ are some given functions of
``momentum'' $p$. Let us express $y_2(p)$ in terms of $\alpha $
and a ``renormalized coupling constant'' $C_0( \mu )\equiv
y_1(p)|_{p=\mu}$. Expressing $C_0$ as
\begin{equation}
C_0={C_0(\mu )\over a(\mu )+\alpha \ b(\mu )} \label{4}
\end{equation}
and substituting eq.~(\ref{4}) into eq.~(\ref{3}), we obtain

\begin{equation}
y_2(p)=\alpha \ d_1(p)+{d_2(p) \ C_0 (\mu )\over a(\mu )+\alpha \
b( \mu )}. \label{5}
\end{equation}

Now, let $\tilde y_2(p)$ denote the result of $y_2(p)$ for $\alpha
=0$. Clearly, $\tilde y_2(p)$ determined from eq. (\ref{3}) as
$y_2(p)$ for $\alpha =0$ does {\it not} depend on $\mu $. On the
other hand, if we naively substitute $\alpha =0$ into eq.
(\ref{5}), we obtain

\begin{equation}
\tilde y_2(p)={d_2(p) \ C_0 (\mu )\over a(\mu )}. \label{7}
\end{equation}
As a consequence, $\tilde y_2(p)$ determined from eq. (\ref{7})
depends on $\mu$, because the $\mu $ dependence of $C_0(\mu )$ is
{\it not} cancelled by $\mu $-dependence of $a(\mu )$:
\begin{equation}
{C_0(\mu )\over a(\mu )}={C_0 \left\{ a(\mu )+ \alpha b(\mu )
\right\} \over a(\mu )}.
\end{equation}

For this simple toy example the resolution of the seeming puzzle
is clear: defining $\tilde y_2(p)$ in terms of ``renormalized
running coupling'', we should substitute $\alpha =0$ in eq.
(\ref{5}) and also replace $C_0(\mu )$ by $\tilde C_0(\mu )$,
where $\tilde C_0(\mu )= y_1(\mu )$ for $\alpha =0$. Doing so we
obtain for $\tilde y_2(p)$

\begin{equation}
\tilde y_2(p)={d_2(p)\tilde C_0(\mu )\over a(\mu )}. \label{10}
\end{equation}
As $\tilde C_0(\mu )=a(\mu ) \ C_0$, eq. (\ref{10}) indeed gives
$\tilde y_2(p)$ which (correctly) does not depend on $\mu $.

The problem of the $\mu $ dependence of the $pp$-scattering length
is fixed in analogy to this toy model. Note that $\tilde C_0(\mu
)$ is uniquely determined by the ``fundamental theory'' and can be
calculated in this  toy model. In EFT $\tilde C_0(\mu )$ is again
uniquely determined by the underlying theory but in practice it is
not possible to calculate it (at least for the moment being) and
therefore has to be given as an input.

%{\bf ACKNOWLEDGEMENTS}

\end{document}